\documentclass[useAMS,usenatbib,12pt]{mn2e}
\usepackage[dvipdf]{graphicx,color}
\usepackage[latin1]{inputenc}
\usepackage{graphics}
\usepackage{amsfonts}
\usepackage{amsmath}
\usepackage{multicol}
\usepackage{layout}
\usepackage{amssymb}
\usepackage{epsfig}
\usepackage[a4paper]{hyperref}
\DeclareGraphicsExtensions{.eps}

\title[Moving barrier merger trees]
{Merger history trees of dark matter haloes in moving barrier models}
\author[J. Moreno et. al.]
       {Jorge Moreno$^{1}$, Carlo Giocoli$^{2}$ \& Ravi K. Sheth$^{1}$ 
\thanks{Email:
\href{mailto:jmoreno,shethrk@physics.upenn.edu}{jmoreno,shethrk@physics.upenn.edu},\,
\href{mailto:carlogiocoli@unipd.it}{carlo.giocoli@unipd.it}}\\
$^{1}$Department of Physics and Astronomy, University of Pennsylvania, 209 South 33rd Street, Philadelphia, PA 19104-6396, USA\\
$^{2}$Dipartimento di Astronomia, Universita degli Studi di Padova, Vicolo dell'osservatorio 2 I-35122 Padova, Italy}
\begin{document}
\date{}
\pagerange{\pageref{firstpage}--
\pageref{lastpage}} \pubyear{2008}
\maketitle
\label{firstpage}
%%%%%%%%%%%%%%%%%%%%%%%%%%%%%%%%%%%%%%%%%%%%%
%             A B S T R A C T               %
%%%%%%%%%%%%%%%%%%%%%%%%%%%%%%%%%%%%%%%%%%%%%
\begin{abstract}
We present an algorithm for generating merger histories of dark matter haloes.  The algorithm is based on the excursion set approach with moving barriers whose shape is motivated by the ellipsoidal collapse model of halo formation.  In contrast to most other merger-tree algorithms, ours takes discrete steps in mass rather than time.  This allows us to quantify effects which arise from the fact that outputs from numerical simulations are usually in discrete time bins.  In addition, it suggests a natural set of scaling variables for describing the abundance of halo progenitors; this scaling is not as general as that associated with a spherical collapse.  We test our algorithm by comparing its predictions with measurements in numerical simulations.  The progenitor mass fractions and mass functions are in good agreement, as is the predicted scaling law.  We also test the formation-redshift distribution, the mass distribution at formation, and the redshift distribution of the most recent major merger; all are in reasonable agreement with N-body simulation data, over a broad range of masses and redshifts.  Finally, we study the effects of sampling in discrete time snapshots.  In all cases, the improvement over algorithms based on the spherical collapse assumption is significant.  
\end{abstract}


%%%%%%%%%%%%%%%%%%%%%%%%%%%%%%%%%%%%%%%%%%%%%
%       I N T R O D U C T I O N             %
%%%%%%%%%%%%%%%%%%%%%%%%%%%%%%%%%%%%%%%%%%%%%
\section{Introduction}
Most models of galaxy formation in a hierarchical universe assume that the merger history of the surrounding dark matter halo plays an important role in determining the properties of a galaxy (e.g. \cite{wr78,wf91,baugh06} and references therein).  Although halo merger histories can be measured using N-body simulations, these can be time consuming and computationally intensive \citep{millennium05}.  This has fueled considerable study of the formation and merger histories of dark matter haloes from a Monte Carlo perspective.  Monte Carlo merger trees have the advantage of being fast and one may easily probe mass regimes inaccessible to current N-body simulations.  Moreover, unlike N-body experiments, the cosmology and initial conditions may be easily modified.  

The excursion set framework \citep{betal91,lc93}, which is motivated by the pioneering work of \cite{ps74}, provides the basis for current models of halo assembly.  Initially, this framework was based on the assumption that haloes form from a spherical collapse, of the type first described by \cite{gg72}.  Fast algorithms for generating halo merger trees, in which haloes were assumed to form from a spherical collapse, were developed in the 1990s \citep{kw93,sk99,sl99,galform00} (see \cite{zmf08} for a review).  However, spherical collapse overpredicts (underpredicts) the abundance of haloes in the low (high) mass regime.  To address these issues, \cite{st99} extended the excursion set framework to include ellipsoidal collapse \citep{smt01,st02}.  This clearly showed that merger-trees which assume spherical collapse are inadequate.  

\cite{hp06} describe a merger tree algorithm which extends some of the older algorithms to incorporate aspects of the ellipsoidal collapse results.  In addition, a number of new algorithms have recently been published \citep{pch07,eyal1}; although efficient and accurate, such methods side-step the idea of ellipsoidal collapse altogether.  Moreover, these methods are callibrated to match N-body simulations, and are therefore limited by the accuracy and scope of these simulations.  

The aim of the present work is to provide a merger history tree algorithm which is based explicitly on the excursion-set formalism with ellipsoidal collapse.  The most significant difference between the algorithm we derive and all the others described above is that it takes discrete steps in mass rather than time.  This feature allows us to study a few problems which are more difficult to address with the other methods.  

After we completed this project, \cite{zmf08} presented an alternative algorithm with ellipsoidal collapse (and discrete-time snapshots).  This algorithm generates progenitors across many mass bins and then assigns them to final haloes.  In this sense, it is very similar to the \cite{kw93} merger tree, but solves many of its problems.  Although this tree is quite different from ours, it also attempts to describe the merger history of a halo from the excursion-set formalism.  These two approaches show that generating merger trees without tuning to N-body simulations is a quite non-trivial, yet interesting, challenge.

A review of background material and a description of our algorithm are given in Section~\ref{tree}.  Section~\ref{nbody} compares our algorithm with excursion-set theory predictions, and with measurements in N-body simulations.  The tests include the progenitor mass fractions and mass functions, the formation-redshift distribution, the mass distribution at formation, and the redshift distribution of the most recent major merger.  Section~\ref{conclusions} summarises our findings and discusses possible applications of our algorithm, an outline of which is in Appendix~\ref{algorithm}.

%%%%%%%%%%%%%%%%%%%%%%%%%%%%%%%%%%%%%%%%%
%         B A C K G R O U N D:          %
%%%%%%%%%%%%%%%%%%%%%%%%%%%%%%%%%%%%%%%%%
\section{Merger trees in the excursion set 
approach}\label{tree}
In the excursion set approach, the problem of estimating halo abundances is mapped to one of estimating the distribution of the number of steps a Brownian-motion random walk must take before it first crosses a barrier of specified height \citep{betal91}.  In this approach, the height of the barrier plays a crucial role.  

%%%%%%%%%%%%%%%%
%   BARRIERS   %
%%%%%%%%%%%%%%%%
\subsection{Constant and moving barriers}
The Press-Schechter mass function is associated with barriers of constant height - such barriers arise naturally in models in which haloes form from a spherical collapse model.  In constrast, in the ellipsoidal collapse model, barriers of the form 
\begin{equation}
\label{ellbarrier}
 B(S,\delta_{c}) =
 \sqrt{q}\delta_{c}\,
 \left\{ 1+\beta \left[\frac{S}{q \delta_{c}^{2}}\right]^{\gamma}\right\}.
\end{equation} are more natural.  Here $\delta_{\rm c}$ is the overdensity required for spherical collapse - it is a monotonically increasing function of redshift, and it is given by $\delta_{\rm c0}/D(z)$, where $\delta_{\rm c0} \equiv \delta_{\rm c}(z=0) \sim 1.686$ and $D(z)$ is the linear growth factor.  $S$ is a monotonically decreasing function of halo mass, given by:
\begin{equation}
 S(m) \equiv \sigma^{2}(m) 
      = \frac{1}{2\pi^2}\,\int {\rm d}k\,k^2\, P(k)\,\tilde{W}^2(kR),
\label{sofm}
\end{equation}
where $P(k)$ is the initial power spectrum of density fluctuations, linearly evolved to the present time, $\tilde{W}$ is the Fourier transform of $W(x) = (3/x^3) (\sin x -x\cos x)$, $R = (3m/4\pi\bar\rho)^{1/3}$, and $\bar\rho$ is the comoving background density.  At large $R$, the overdensity contained in the associated volume is practically zero.  As $R$ (and $m(R)$) decreases, $S(R)$ increases, and $\delta_{R}$ executes a random walk.  We refer the reader to \cite{lc93} for more details.

Consider a barrier $B(S,\delta_{\rm c}(z))$, as in equation~(\ref{ellbarrier}), and an ensemble of random walks which start from the origin: $(S,\delta)=(0,0)$.  The excursion set approach maps the distribution of $S$, the number of steps a random walk must take to first cross such a barrier, to the fraction of mass in $m$-haloes at redshift $z$.  This quantity is associated with the so-called unconditional mass function.  The conditional mass function of high redshift progenitors of a more massive final $M$-halo at some lower redshift $Z$ is modeled using walks which start from $(S_M,\delta_{\rm c}(Z))$ instead. 

The shape of the mass functions (unconditional and conditional) is determined by the shape of the barrier, encoded in the $(q,\beta,\gamma)$ parameters.  The spherical collapse model is associated with $(q,\beta,\gamma)=(1,0,0)$, whereas ellipsoidal collapse has $(0.707,0.47,0.615)$.  When $\gamma > 1/2$, not all walks are guaranteed to cross the ellipsoidal collapse barrier \citep{st02}.  Moreover, the barriers associated with two different times may intersect; of course, this never happens for the spherical collapse barriers.  Sheth \& Tormen suggest that this intersection of barriers may represent the possibility that haloes can fragment.  This complicates the algorithm we describe below, so we instead study the limiting case of `square-root' barriers for which $\gamma = 1/2$:  
\begin{equation}
\label{sqrtB}
 B(S,\delta_{c}) = \sqrt{q}\delta_{c} + \beta\sqrt{S}.
\end{equation}  
The predicted halo abundances associated with $(q,\beta,\gamma)=(0.55,0.5,0.5)$ are very similar to those in simulations \citep{mr05,ms08}.  Moreover, the first crossing distribution of this barrier is known \citep{breiman}.

%%%%%%%%%%%%%%%%%%
%  ETA-SYMMETRY  %
%%%%%%%%%%%%%%%%%%
\subsection{A conditional scaling symmetry}\label{wsim}
Recall that the unconditional mass function is associated to the first crossing distribution associated with walks which start from the origin: $(S,\delta)=(0,0)$.  When constant barriers are use, it can be expressed self-similarly as
\begin{equation}
f(m|z) {\rm d}m=f(S|\delta_{\rm c}) {\rm d}S=f(\nu) {\rm d} \nu,
\end{equation}
where $\nu\equiv\delta_{c}^{2}/S$.  The conditional mass function of $m$-haloes at $z$ that end up in bound objects of mass $M>m$ at $Z<z$ is given by $f(m,z|M,Z){\rm d}m=f(S_m,\delta_1|S_M,\delta_0){\rm d}S_m$, where  where $\delta_1=\delta_{\rm c}(z)$, $\delta_0=\delta_{\rm c}(Z)$.  In other words, the conditional mass function is associated with the first crossing distribution of a barrier of height $\delta_1$ by random walks with origin at $(S_M,\delta_0)$.  Because a straight-line is straight whatever the origin of the coordinate system, the conditional mass function, in the spherical collapse model has the same functional form as that of the unconditional mass function, provided one sets $\nu = (\delta_1 - \delta_0)^2/(S_m-S_M)$.  

However, for the square-root barrier, a walk which starts from $(\sqrt{q}\,\delta_0 + \beta \sqrt{S_M},S_M)$ must cross a barrier of shape $\sqrt{q}(\delta_1-\delta_0) + \beta\sqrt{S_m-S_M + S_M}$.  This is not quite of the same form as equation~(\ref{sqrtB}).  As a result, the conditional mass function is not simply a rescaled version of the unconditional one.  Rather, in this model, 
\begin{equation} 
 f(m,z|M,z_{0})\,{\rm d}m = f(S_m/S_M|\eta)\,{\rm d}(S_m/S_M), %\\
 % \approx f(m/M|\,\eta)\,{\rm d}(m/M), 
\end{equation}
where 
\begin{equation}
\label{omega}
 \eta \equiv \frac{\delta_1-\delta_0}{\sqrt{S_M}}.
\end{equation}
(see \cite{breiman} or \cite{gmst07} for the exact expressions).  Thus, final halos of different masses will have similar progenitor mass functions when expressed in terms of $S_m/S_M$, provided they have similar values of $\eta$.  While this scaling is like that for the constant barrier model, in the square-root barrier model, the progenitor mass function is {\em not} a function of the combination $\nu^2=\eta^2/(S_m/S_M - 1)$.  This is interesting, because \cite{st02} have shown that the conditional mass function in simulations is not well-fit by a function of $\nu$.  In what follows, we will present evidence that it is, however, a function of $\eta$ and $S_m/S_M$ separately, so the qualitatively different scaling associated with the square-root barrier is indeed seen in simulations.  

%%%%%%%%%%%%%%
\begin{figure}
\centering
\includegraphics[width=0.95\columnwidth]{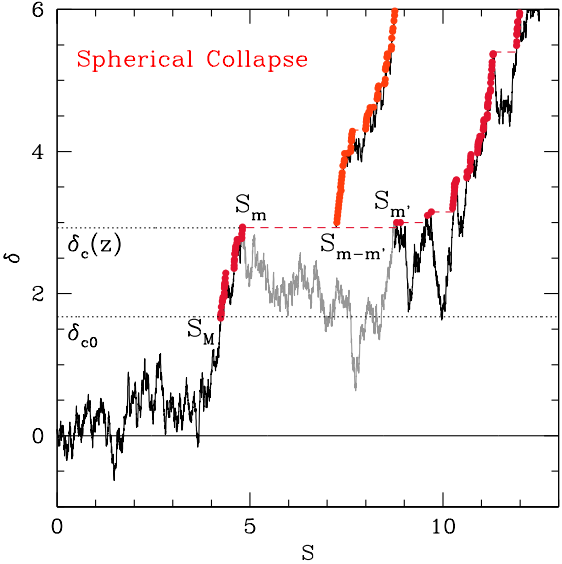}
\caption{A random walk and its associated mass history.  The dark filled circles represent the history of a halo of mass $M$ at redshift $z=0$.  A merger $(m',m-m') \rightarrow m$ at redshift $z$ is depicted by the $S_{m} \rightarrow S_{m'}$ {\it jump} at height $\delta_{c}(z)$.  A new branch associated with $(m-m')$ is connected at $(S_{m-m'},\delta_{c}(z))$.  The light-filled circles denote the mass history of this object.}
\label{chist}
\end{figure}
%%%%%%%%%%%%%%

%%%%%%%%%%%%%%
\begin{figure}
\centering
\includegraphics[width=0.95\columnwidth]{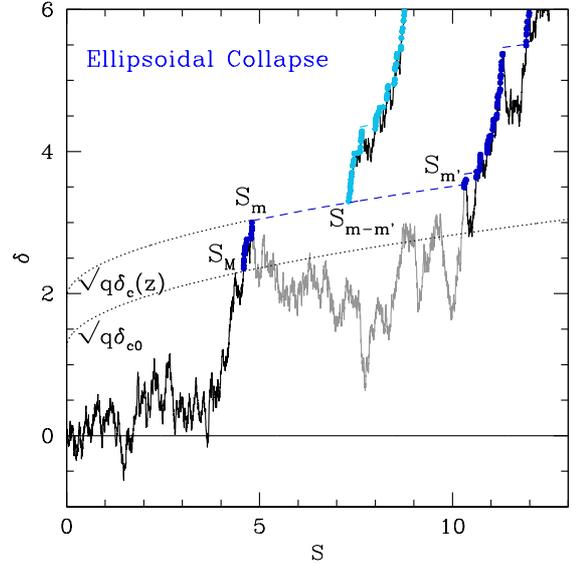}
\caption{The same random walk as in Figure~\ref{chist}, but now with square-root rather than constant barriers, illustrating that the mass accretion history depends on the barrier shape.  In our algorithm, the new object with mass $(m-m')$ is now connected at $(S_{m-m'},\sqrt{q}\delta_{\rm c}(z)+\beta\sqrt{S_{m-m'}})$.}
\label{shist}
\end{figure}
%%%%%%%%%%%%

%%%%%%%%%%%%%%%
\begin{figure*}
\centering
\includegraphics[width=0.9\hsize]{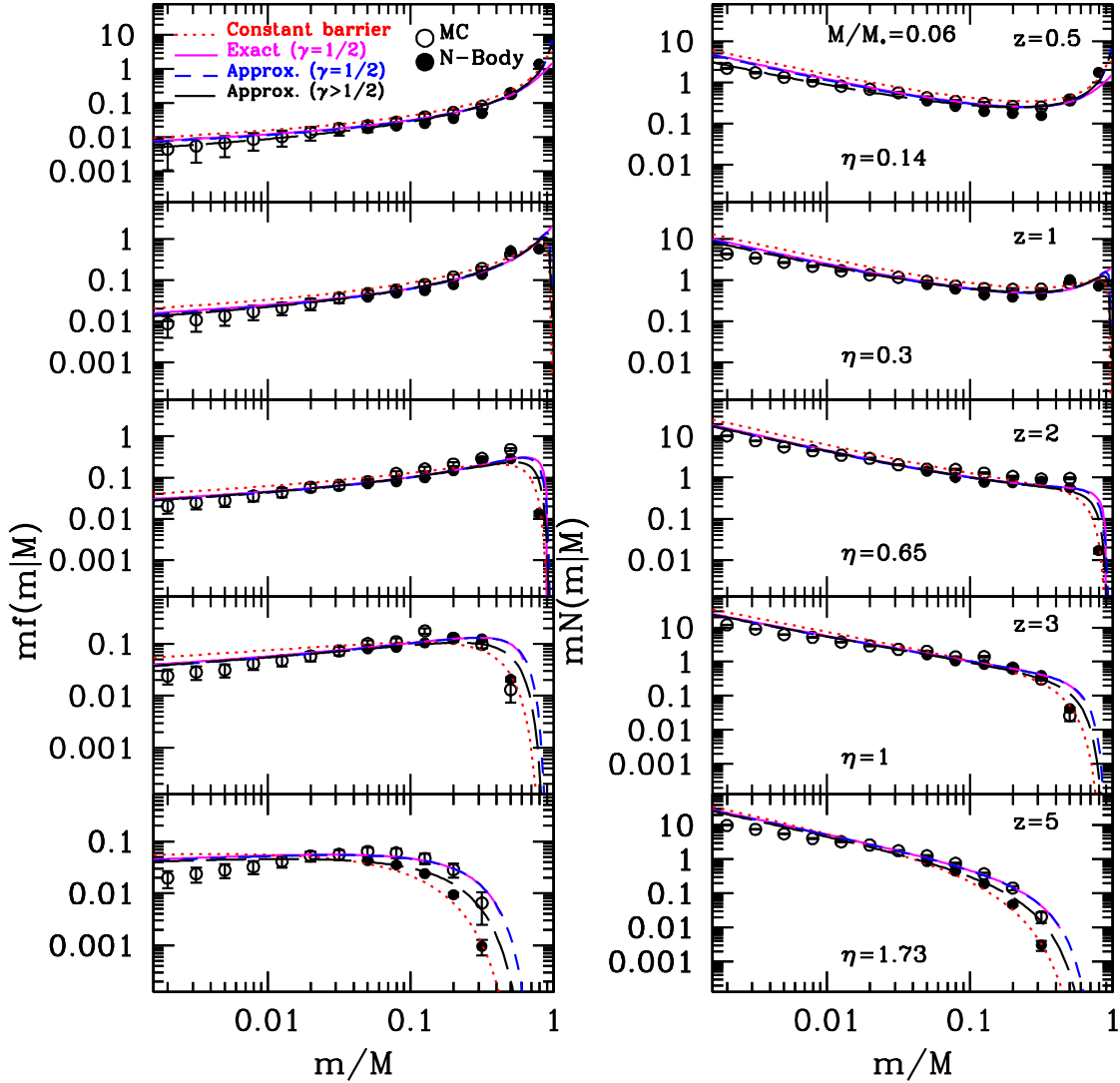}
\caption{The progenitor mass fraction (left) and mass function (right) at redshifts $z=(0.5,1,2,3,5)$, for haloes of mass $M/M{\star}=0.06$ at $z=0$.  Filled circles show measurements in the GIF2 simulation, and open circles are from our square-root trees.  The smooth solid and dashed curves show the exact square-root barrier solution, and the series approximation, respectively.  The long-dashed curve shows the ellipsoidal collapse model with $\gamma>1/2$, and the short-dashed curve is the constant barrier prediction.  Values of the scaling parameter $\eta$ (equation~\ref{omega}) are also shown (see Figure~\ref{wprog}).}
\label{prog1}
\end{figure*}
%%%%%%%%%%%%%

%%%%%%%%%%%%%%%
\begin{figure*}
\centering
\includegraphics[width=0.9\hsize]{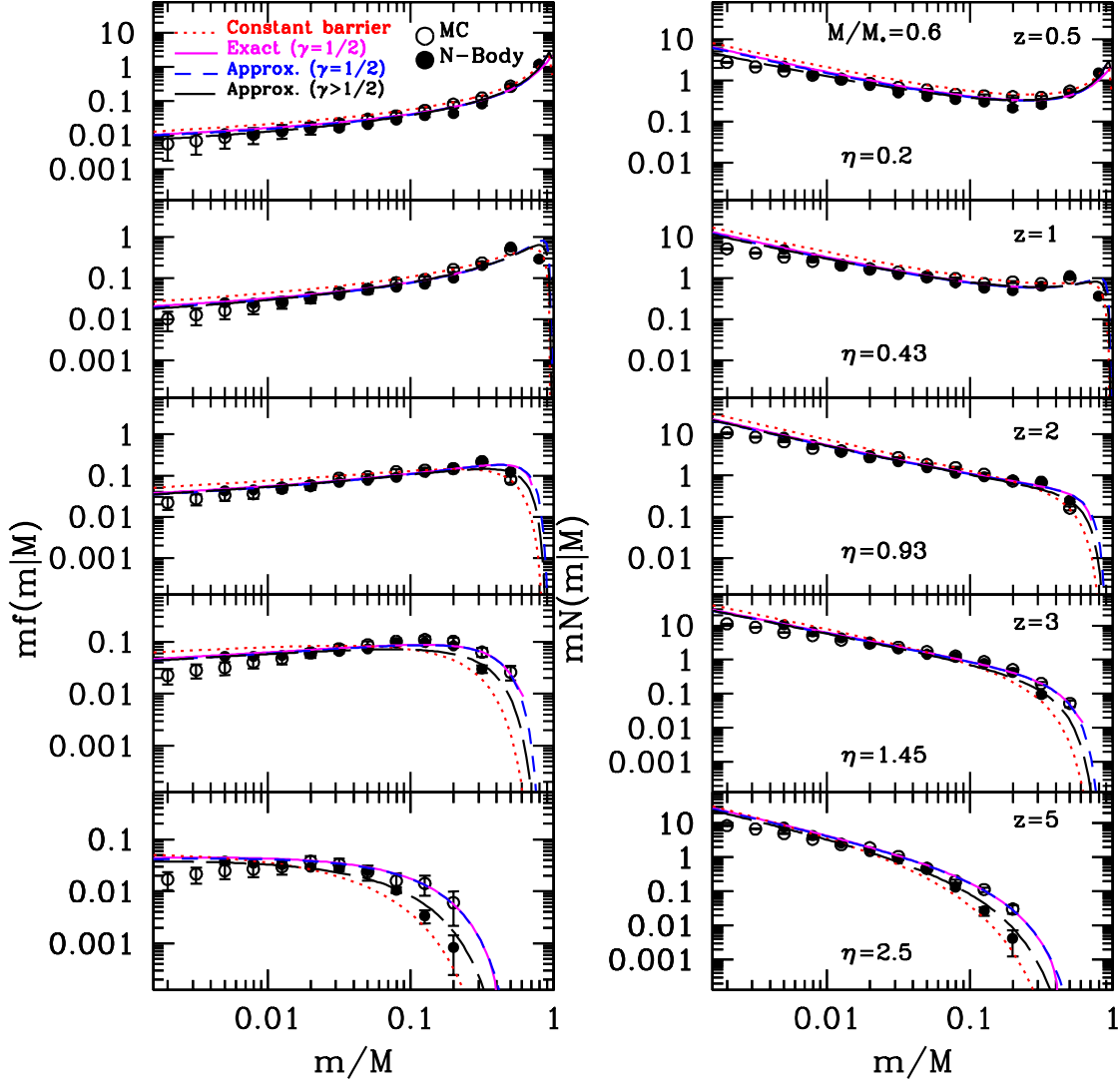}
\caption{Same as Figure~\ref{prog1}, but with
$M/M_{\star}=0.6$.}
\label{prog3}
\end{figure*}
%%%%%%%%%%%%%

%%%%%%%%%%%%%%%
\begin{figure*}
\centering
\includegraphics[width=0.9\hsize]{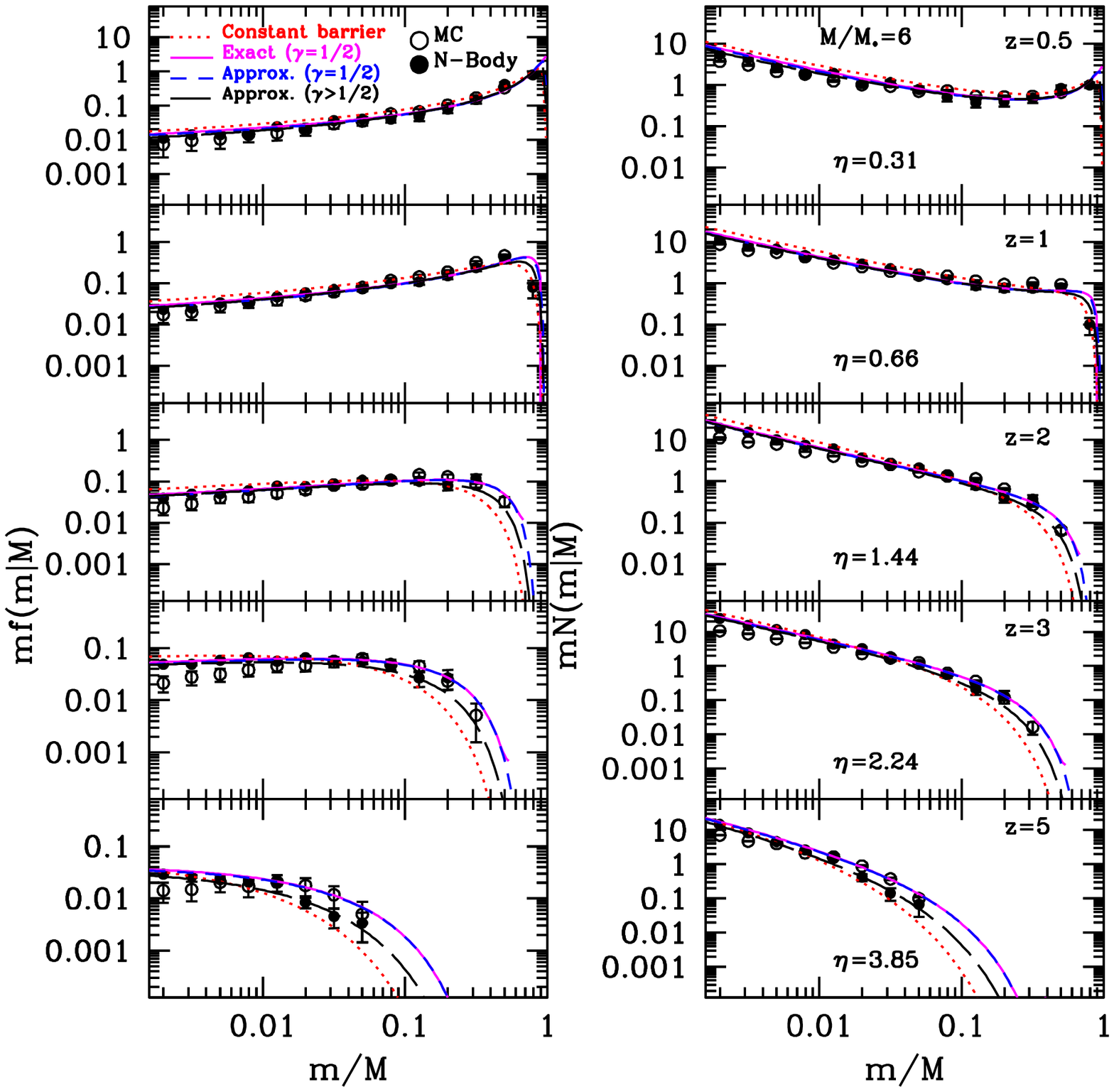}
\caption{Same as Figures~\ref{prog1}, but 
with $M/M_{\star}=6$.}
\label{prog5}
\end{figure*}
%%%%%%%%%%%%%

%%%%%%%%%%%%%%%%%%%%%%%
%  HISTORIES & TREES  %
%%%%%%%%%%%%%%%%%%%%%%%
\subsection{Mass histories and merger trees}\label{masshist}
Figure~\ref{chist} illustrates how the mass growth history of an object is encoded in the excursion set approach if objects form from a spherical collapse (also see Figure~1 of \cite{lc93}).  The jagged line shows a random walk which starts from the origin: $(S,\delta)=(0,0)$.  Imagine drawing a horizontal line with height $\delta_{\rm c0}=1.686$ and marking the {\it smallest} value of $S$ at which the walk intersects this `barrier' of constant height ($\delta_{\rm c0}$ corresponds to the present time and $\delta_{c} > \delta_{\rm c0}$ corresponds to higher redshifts).  The dotted horizontal line denotes such barrier.  Then increase the height of this barrier, and record how this value of $S$ changes as $\delta$ increases.  Such mass history points are depicted as dark filled circles in Figure~\ref{chist}.  The dashed lines show that $S$ will occasionally jump from a small value to a larger one.  Since $S$ is a proxy for mass, and $\delta_{\rm c}$ for time, such a jump is a proxy for an instantaneous change in mass:  a merger.  Note that the random walk steps under such jumps are not part of the mass history (e.g., the gray portion with $S_m < S < S_{m'}$).

The key to our merger tree algorithm, which is described in detail in Appendix~\ref{algorithm}, is to recognize that these jumps mean that there are a set of other walks which one might associate with this one -- one for each jump.  One such walk is illustrated by the second jagged curve, which starts at about the middle of the panel.  If the jump from $S_m$ to $S_{m'}$ occurred when the barrier height was $\delta_{\rm c}(z)$, then this other walk starts from $(S_{m-m'},\delta_{\rm c}(z))$.  The `merger history' associated with this new {\it branch} is represented by the light-shade filled circles in Figure~\ref{chist}.  For every such jump, a new random walk must be drawn.  For each jump within each of those new walks, the same process applies -- more walks must be drawn.  In summary, the bundle of such walks encodes the entire merger history of a present-day object.  Notice that jumps can occur at any $z$ -- there is no constraint that they happen at discrete times.  However, if one is interested in the mass function of progenitors at some fixed $z$, one simply reads-off the list of values of $S$ at which this bundle of walks first cross $\delta_{\rm c}(z)$.  

So far, we have discussed how to generate trees in the spherical collapse model.  Figure~\ref{shist} shows the same walk as before, but now the mass growth history associated with the walk is given by its intersection with square-root barriers of gradually increasing height.  This shows clearly that the jumps in mass, and the times at which they occur, are modified.  But the overall logic remains the same.  Each jump gives rise to a new walk that starts from $(S_{m-m'},B(S_{m-m'},\delta_{\rm c}(z)))$, where $B$ is given by equation~(\ref{sqrtB}).

The natural generalisation to spherical collapse is to incorporate the original $\gamma>1/2$ barrier of \cite{smt01}.  Such a choice would complicate this algorithm significantly.  First of all, as Figure~\ref{shist} illustrates, the shape of the square-root barrier remains the same with different redshifts, except for an overall vertical shift.  This is not the case with $\gamma>1/2$.  As $\delta_{\rm c}$ increases (increasing redshift), the term $\delta_{\rm c}^{1/2-2\gamma}$ makes the barrier in equation~(\ref{ellbarrier}) increase less rapidly with $S$.  A consequence of this is that the barriers associated with different redshifts cross.  In the absence of crossing-barriers (e.g., constant and square-root barriers), one may uniquely map any point in the $(S,\delta)$ plane to $(m,z)$.  The crossing of barriers invalidates this property, making the identification of jumps with mergers at a given time ill-defined.

%%%%%%%%%%%%%%%%%%%%%%%%%%%%%%%%%%%%%%%%%%%%
% C O M P A R E   W/  S I M U L A T I O N  %
%%%%%%%%%%%%%%%%%%%%%%%%%%%%%%%%%%%%%%%%%%%%
\section{Comparison with simulations}
\label{nbody}
In this section, we compare the statistical properties of our merger history trees with expectations from the excursion set theory which they are supposed to reproduce, and with measurements in the GIF2\footnote{German Israel Fund.} N-body simulation \citep{gao04b}.  The simulation followed the evolution of $400^3$ particles in a periodic cubic box $110h^{-1}$Mpc on a side in a flat ${\rm \Lambda}$CDM background cosmology with parameters ($\Omega_m,\sigma_8,h,\Omega_bh^2,n) = (0.3,0.9,0.7,0.0196,1)$.  Fifty simulation snapshots were output, equally spaced in log$(1+z)$.  At each snapshot, haloes were identified using the spherical overdensity criterion, adopting for virial mass the definition of \cite{eke} (i.e., with virial density at $\sim 324\bar{\rho}$ at redshift zero).  The particle mass is $m_{\rm p} = 1.73 \times 10^{9} h^{-1} {\rm M_\odot}$ and only objects with at least ten particles are considered.  $M_{\star}(z)$, defined by $\delta^{2}_{c}(z)=S(M_{\star}(z))$, is the typical mass scale at redshift $z$.  It is common practice to express halo masses in terms of $M_{\star}=M_{\star}(z=0)$ (the $z$-dependence is suppressed for the present time).  For this cosmology and initial power spectrum, $M_{\star}= 8.7 \times 10^{12}h^{-1} {\rm M_{\odot}} \simeq 5030 m_{\rm p}$.  The simulation data and halo catalogues are available at \href{http://www.mpa-garching.mpg.de/Virgo}{\texttt{http://www.mpa-garching.mpg.de/Virgo}}.  See \cite{getal08} for more details regarding the post-processing of the simulation.

To compare our merger histories with those in the GIF2 simulation, we generated 2000 realisations of our tree for each final halo mass bin $M$ of interest.  In all cases, the minimum mass considered was $m_{\rm dust}=M/1000$, and the merger histories of haloes with mass below this were not followed (we call this minimum mass the `branching-mass resolution').  We used random walks with $10^{5}$ steps in between $S_{M}$ and $S_{\rm dust}$ to ensure that the mass change between the steps was less that $m_{\rm dust}$.  Having a small step size is essential to faithfully reproduce random walks in a computer.  Moreover, if the step size is too large, we run the risk of missing branches.  Numerical tests showed that the outputs converged to our results for small step sizes.  See the Appendix for more details on the implementation of our Monte Carlo tree. Recall that our tree does not take discrete steps in time.  Nevertheless, for fair comparison with the measurements from the GIF2 simulation, the tree data were stored in the same discrete redshift bins as were output from the simulation.  We use `Cont' to denote the original tree data and `Snap' for the data stored in redshift snapshots.  Sections~\ref{mformation} and \ref{lastmerger} study some merger-related quantities which are sensitive to the differences between these two ways of storing trees (Figures~\ref{mF}, \ref{zoom} and \ref{zlmm}).  

%%%%%%%%%%%%%%%
\begin{figure*}
\centering
\includegraphics[width=0.9\hsize]{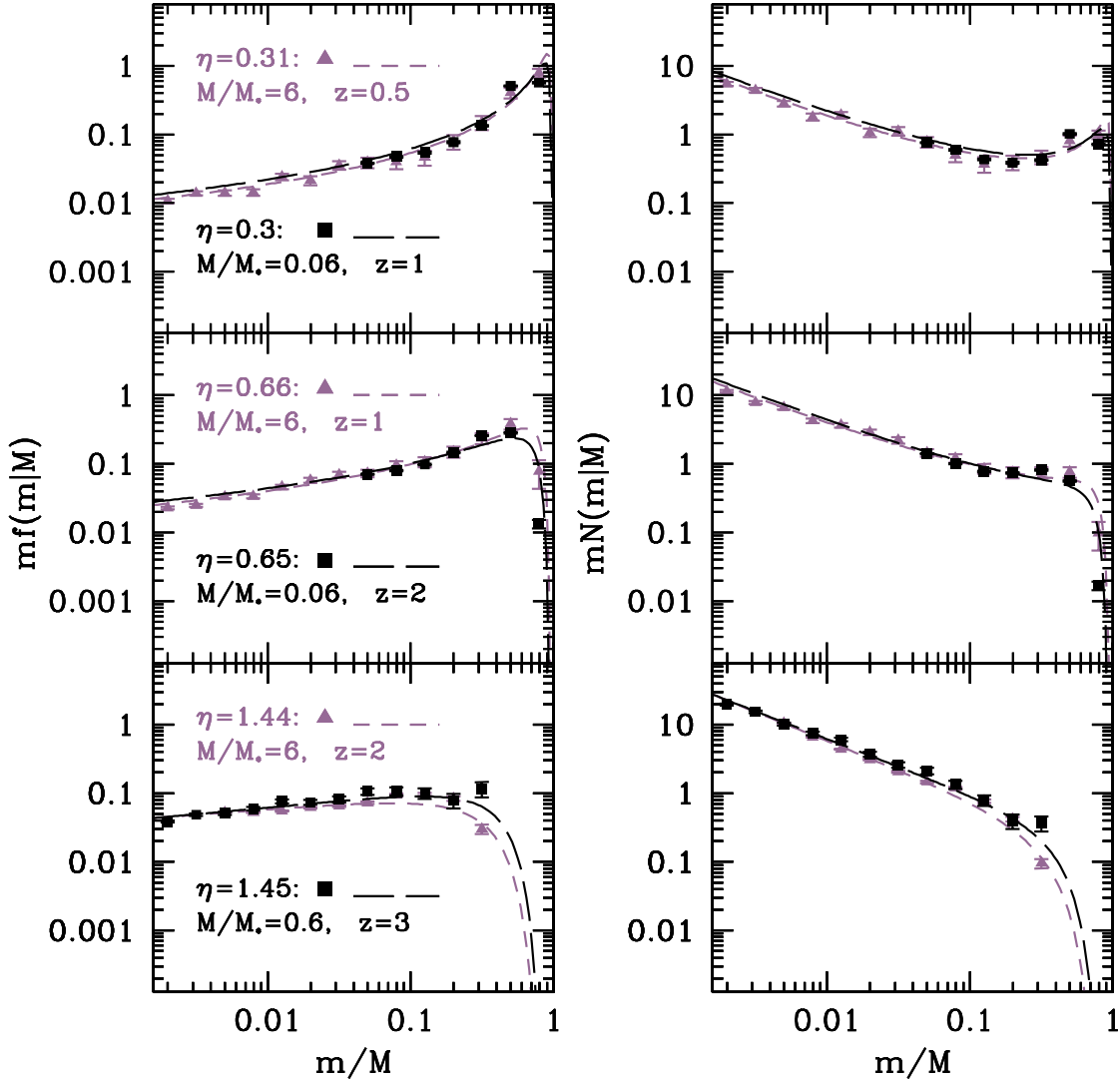}
\caption{The $\eta$-symmetry.  Different combinations of $M$ and $z$ with similar $\eta$ (equation~\ref{omega} and Table~\ref{wtable}).  N-body simulation measurements and the \citet{st02} result with $\gamma > 1/2$ are shown.}
\label{wprog}
\end{figure*}
%%%%%%%%%%%%%

%%%%%%%%%%%%%%%%%%%%%%%%%%%%%%
%  PROGENITOR MASS FUNCTION  %
%%%%%%%%%%%%%%%%%%%%%%%%%%%%%%
\subsection{The progenitor mass function}
\label{prog}

Figures~\ref{prog1}-\ref{prog5} show the 
progenitor mass fractions and mass functions 
at five different redshifts ($z=0.5,1,2,3,5$), for haloes identified at $z=0$ with final masses given by $M/M_{\star}=0.06,0.6$ and 6.  The corresponding values of $\eta$ (equation~\ref{omega}) are shown in each panel.  In all three figures, filled circles show measurements in the GIF2 simulation, and open circles show results from our square-root trees.  We probe the $m<m_{\rm p}$ regime with our trees to verify consistency with analytic excursion-set predictions (the smooth curves in all the panels).  The expressions we use are given explicitly in Appendix~A of \cite{gmst07}.  The short-dashed curve shows the constant barrier  $(\beta,q)=(0,1)$ prediction associated with spherical collapse \citep{lc93}.  The solid curves show the exact square-root barrier solution with $(\beta,q,\gamma)=(0.5,0.55,0.5)$; this is a complicated affair, involving sums of Parabolic Cylinder functions \citep{breiman}.  Dashed curves show the considerably simpler approximation to the solution which is due to \cite{st02}; this approximation is excellent over the entire range of interest.  The long-dashed curves show this same approximation for the ellipsoidal collapse barrier:  $(\beta,q,\gamma)=(0.707,0.47,0.615)$.  The square-root barrier prediction agrees well with the $\gamma=0.615$ curve, except in the high-mass regime.  This discrepancy becomes evident when $\eta>1$ and it is amplified with increasing $\eta$ (see below).

Before we ask how our merger tree algorithm compares with simulations, we note that it produces progenitor mass functions that are well-described by the theory curves over a wide range of masses and redshifts.  At high redshifts, our tree data lie slightly below the theory curves at both high and low $m/M$, and slightly above in between, although where the cross-over points occur depends on $z$ and $M$.  In all other regimes, our Monte Carlo trees match the square-root barrier predictions.  Any additional disagreement with the GIF 2 simulation measurements (compare open and filled circles) is due to limitations of the $\gamma=1/2$ model.

%%%%%%%%%%%%%%%
\begin{figure*}
\centering
\includegraphics[width=0.9\hsize]{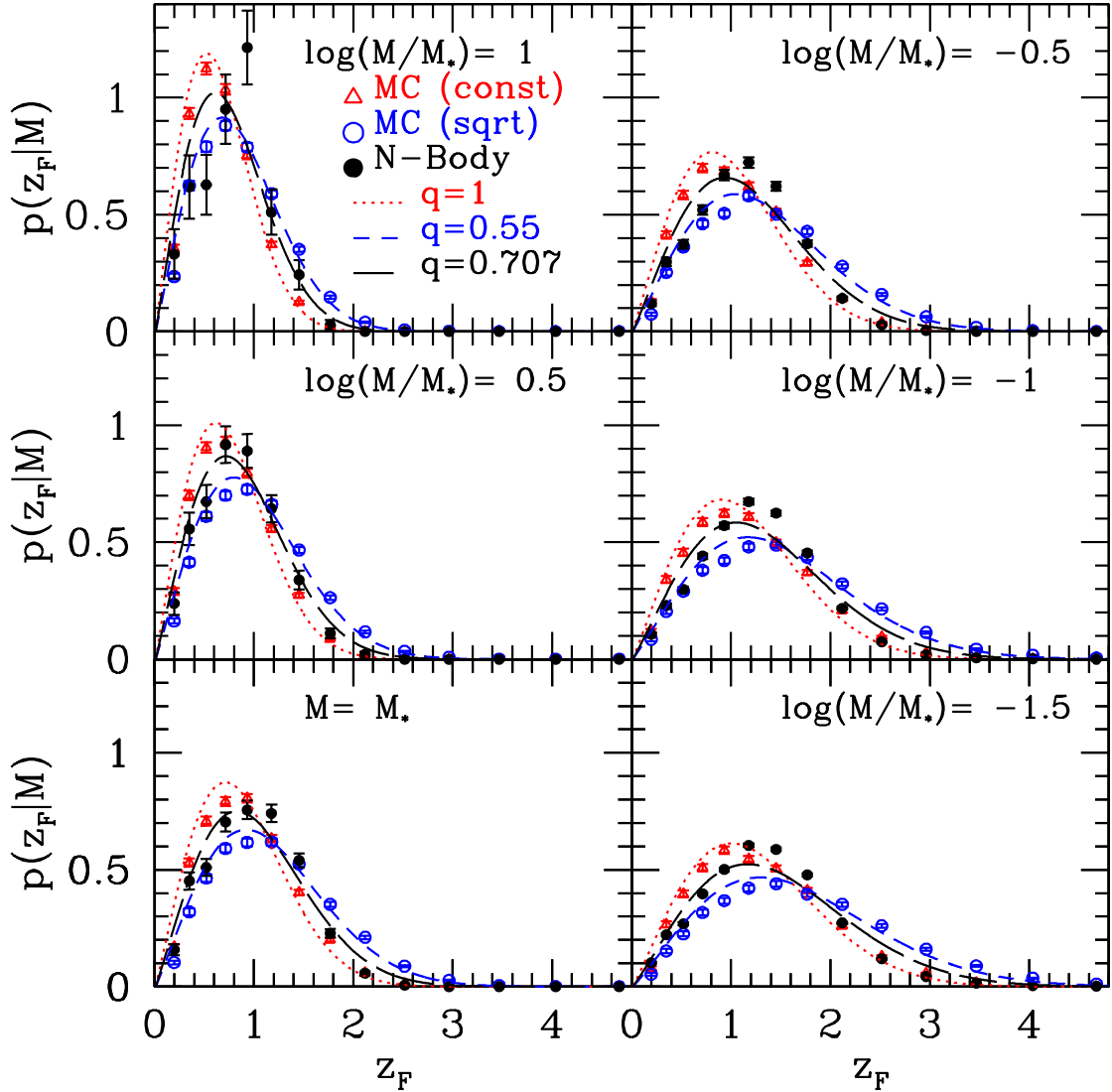}
\caption{Distribution of formation redshifts.  Filled circles show simulation data, open circles and triangles show results from the square-root and constant barrier trees.  Smooth curves show equation~(\ref{pw}) with $q=1$, 0.707 and 0.55 (short-dashed, long-dashed, and dashed), corresponding to the predicted distribution for constant barriers (spherical collapse), moving (ellipsoidal collapse) and square-root barriers.}
\label{zF}
\end{figure*}
%%%%%%%%%%%%%

%%%%%%%%%%%%%%%
\begin{figure*}
\centering
\includegraphics[width=0.9\hsize]{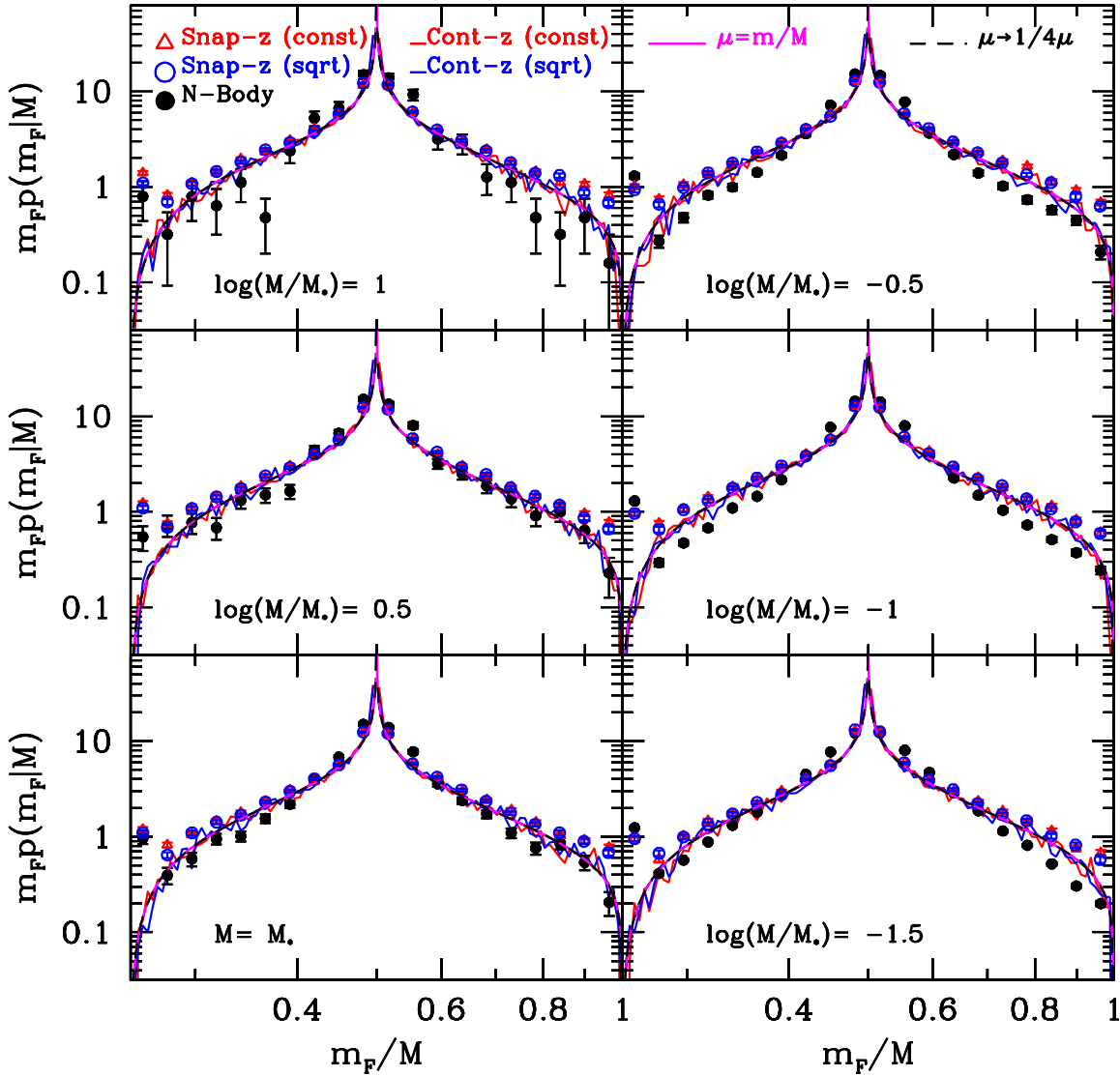}
\caption{Distribution of the mass at formation for several final masses.  The left half of each panel shows the mass just prior to formation, whereas the right half shows the mass just after formation.  Filled circles show simulation data, open circles and triangles are from the square-root and constant barrier trees.  The solid curve shows $\mu q(\mu)$ (right half) and $\mu p(\mu)$ (left half) (equations~\ref{qu} and~\ref{pu} respectively).  The dashed curve shows these same expressions with 
$\mu \rightarrow 1/4\mu$.}  
 \label{mF}
\end{figure*}
%%%%%%%%%%%%%

%%%%%%%%%%%%%%
\begin{figure*}
\centering
\includegraphics[width=0.9\hsize]{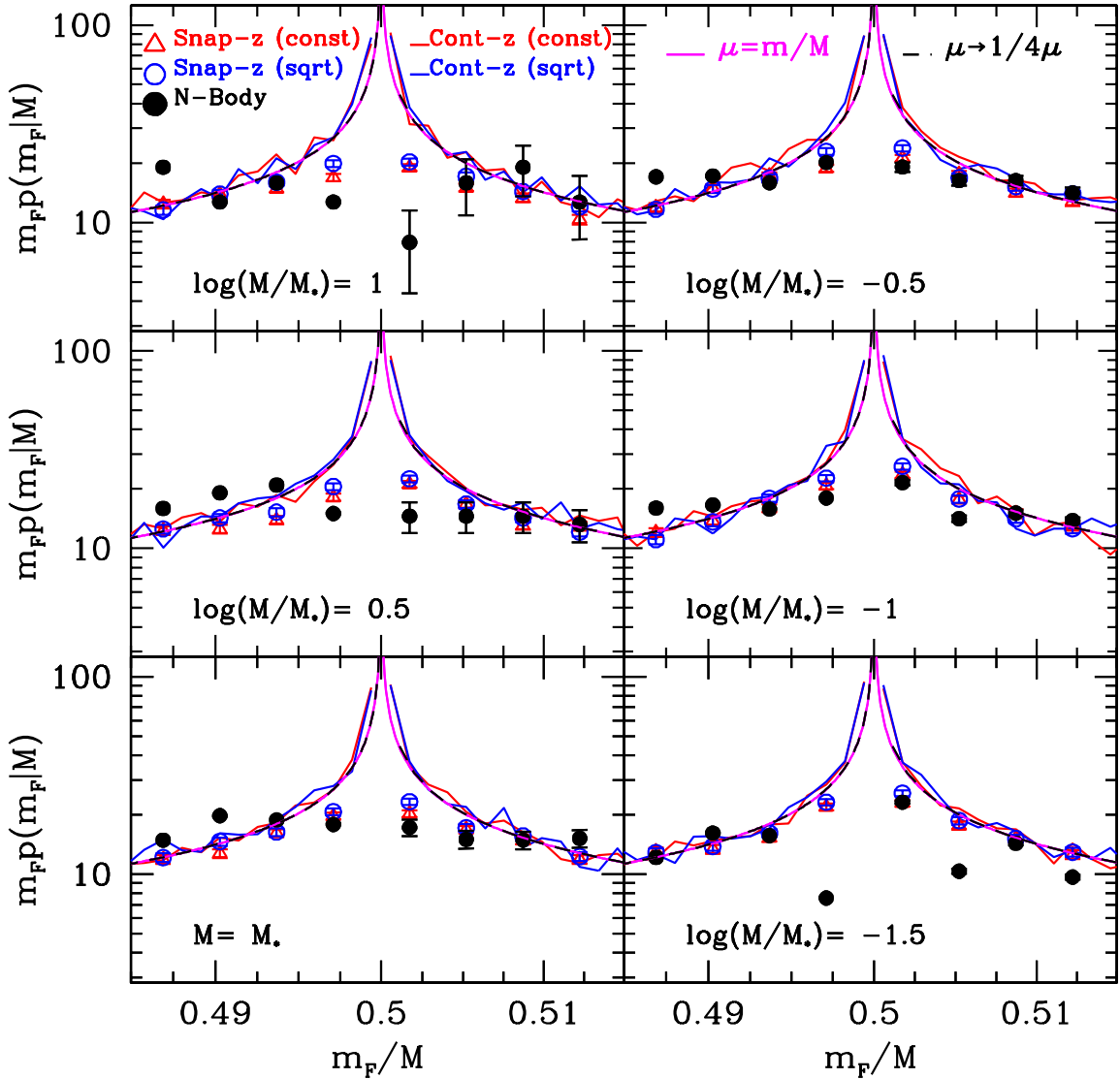}
\caption{Same as Figure~\ref{mF}, but showing only the region around $m/M=1/2$.  The peak in the simulations (filled symbols) is less pronounced than in the merger trees (jagged lines).  Open circles show the result of sampling the merger trees at the same redshifts as the simulation snapshots: this makes a dramatic difference around $0.49\le \mu\le 0.51$, suggesting that the sharp cusp predicted by the theory will also be present in simulations with sufficiently closely spaced outputs. The smaller discrepancies further from the peak remain.}
\label{zoom}
\end{figure*}
%%%%%%%%%%%%

%%%%%%%%%%%%%%%
\begin{figure*}
\centering
\includegraphics[width=0.9\hsize]{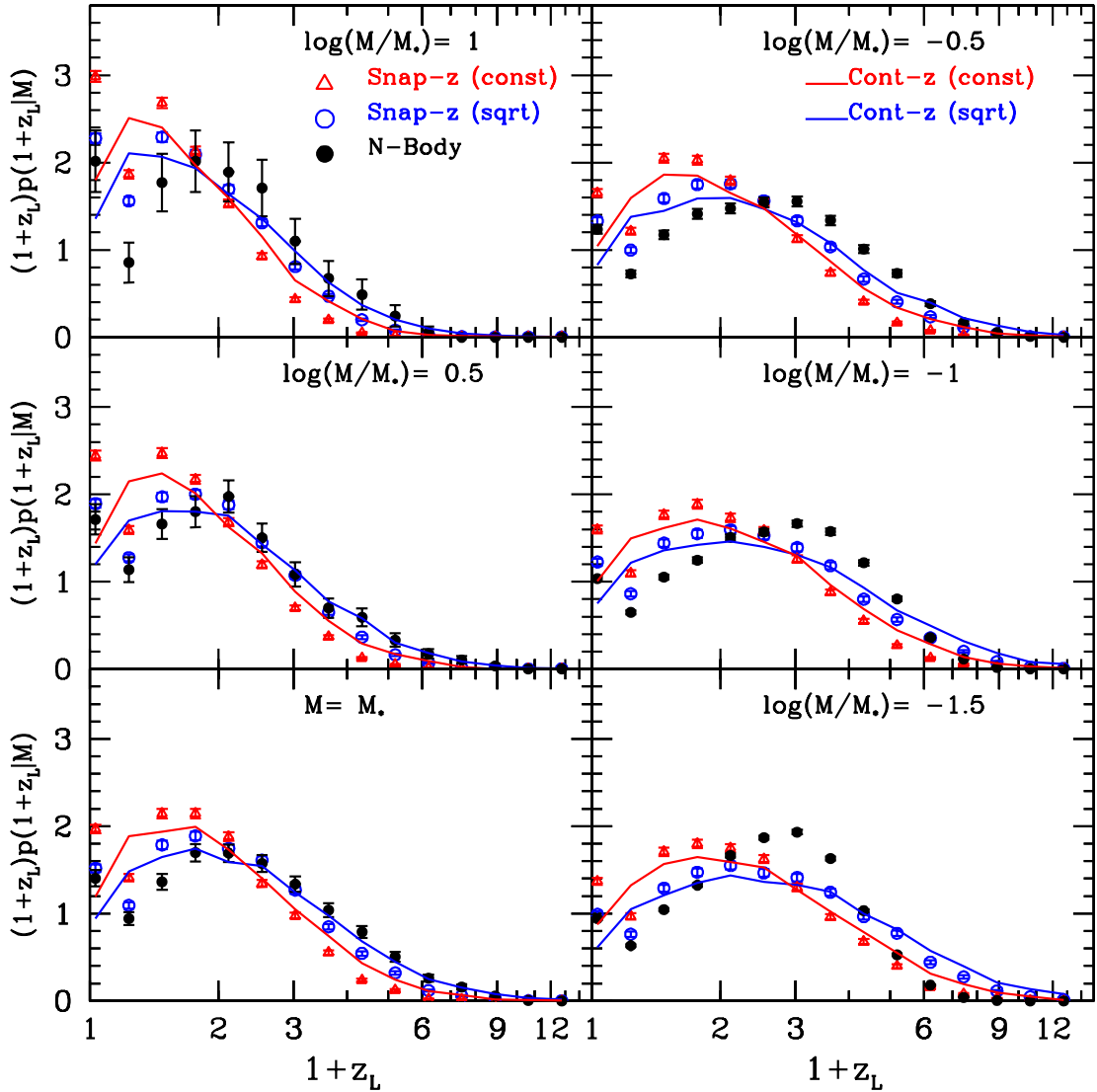}
\caption{The redshift distribution of the most recent major merger, for several final masses.  A major merger is defined as one in which the minor component has at least 1/3 of the mass of the major component.  Filled circles show the GIF2 simulation data, open symbols show the corresponding measurements in the same `snapshot' versions of our trees, and smooth curves show the `continuous' distributions which would be seen with arbitrarily closely spaced output times.}
\label{zlmm}
\end{figure*}
%%%%%%%%%%%%%

%%%%%%%%%%%%%
\begin{table}
\label{wtable}
\begin{center}
\begin{tabular}{ l | l | l | c | l | l | l }  \hline
  $M/M_{\star}$ & $z$ & $\eta$ & | & $M/M_{\star}$ & $z$ & $\eta$ \\ \hline \hline
  0.06 & 1 & 0.3 & | & 6 & 0.5 & 0.31  \\ \hline 
  0.06 & 2 & 0.65  & | & 6 & 1 & 0.66  \\ \hline 
  0.6 & 3 & 1.45  & | & 6 & 2 & 1.44  \\ \hline  
\end{tabular}
\caption{The $\eta$-symmetry (equation~\ref{omega}) used for the comparisons shown in Figure~\ref{wprog}.}  
\end{center}
\end{table}
%%%%%%%%%%%

Finally, recall that the square-root and constant barrier models make specific predictions for how the conditional mass functions should scale with final halo mass and time.  Table~\ref{wtable} lists pairs with similar $\eta$, yet quite distinct values of $M$ and $z$.  Figure~\ref{wprog} compares the associated conditional mass functions.  The black squares and long-dashed lines refer to the left-hand side of Table~\ref{wtable} (low final masses), whereas the gray triangles and short-dashed lines refer to the right-hand side (high final masses).  Notice that the curves are remarkably similar to one another, as are the symbols.  This is true despite the fact that the values of $\eta$ are not perfectly identical, and that $f(m,z|M,Z){\rm d}m = f(S_m/S_M|\eta){\rm d}(S_m/S_M) \simeq f(m/M|\eta){\rm d}(m/M)$.  The results for low-mass haloes (black squares) are truncated at higher $m/M$ than they are for larger $M$ (gray triangles), simply because only haloes with at least ten particles are considered.  Evidently, the conditional mass functions are indeed functions of $\eta$ and $S_m/S_M$ separately, rather than of the combination $\nu$.  

%%%%%%%%%%%%%%%%%%%%%%%%
%  FORMATION REDSHIFT  %
%%%%%%%%%%%%%%%%%%%%%%%%
\subsection{The distribution of formation redshifts}\label{zformation}
Following \cite{lc93}, a halo is said to have `formed' when it first acquires half of its final mass.  For a given halo mass, there is a distribution of formation redshifts.  This distribution is expected to peak at earlier times for lower mass haloes.  The excursion set model with constant barriers provides a simple expression for this distribution of formation times: 
\begin{equation}
 p(\omega_{F})\, {\rm d}\omega_{F} = 
  2\omega_{F}\,{\rm erfc}(\omega_{F}/\sqrt{2})\,{\rm d}\omega_{F},
 \label{pw}
\end{equation}
where 
\begin{equation}
 \omega_{F} \equiv \frac{\delta_{\rm c}(z_{\rm F})-\delta_{\rm c}(z_0)}                          {\sqrt{S(M/2)-S(M)}}
            = \frac{\eta}{\sqrt{S(M/2)/S(M)-1}},
\label{omegaf}
\end{equation}
with $\eta$ given by equation (\ref{omega}).

The filled circles in Figure~\ref{zF} show the formation redshift distributions for haloes with masses in the range $0.9M \leq M \leq 1.1M$ with $\log_{10}(M/M_*) = \{1, \dots, -1.5 \}$ in steps of $-0.5$ in the GIF2 simulation.  We use the first snapshot when at least half the mass is in a single progenitor as the formation time, and make no attempt to 
interpolate our simulation formation redshifts between these discretely spaced output times \citep{harker,gmst07}.  The open circles and triangles show the corresponding formation time distributions from our square-root and constant barrier trees.  Recall that we do not discretise redshift in our tree, so the question of interpolation does not arise. 

For the ellipsoidal collapse model, \cite{gmst07} showed that the formation redshift is well-described by equation~(\ref{pw}) if one replaces $\omega_F \rightarrow \sqrt{q}\omega_F$.  The smooth curves show this with $q=1$, 0.707 and 0.55 (short-dashed, long-dashed, and dashed), which represent the (constant, $\gamma=0.615$, and square-root) barrier predictions.  For higher values of $q$, the peaks are located at lower redshifts and the widths of the curves decrease.  Strictly speaking, equation~(\ref{pw}) only holds for white-noise initial conditions.  Nevertheless, as pointed out \cite{lc93}, it remains a reasonable approximation to CDM case.  Furthermore, note that it provides an excellent description of the formation times generated by our trees.  However, no choice of $q$ provides particularly good agreement with the GIF2 simulation, a discrepancy noted by previous authors \citep{linjinglin,hp06,gmst07}.  This is likely a consequence of the excursion set assumption that different steps in the walk are uncorrelated \citep{st02}.  See \cite{pan} and references therein for how one might improve on this.

%%%%%%%%%%%%%%%%%%%%%%%
%  MASS AT FORMATION  % 
%%%%%%%%%%%%%%%%%%%%%%%
\subsection{The mass distribution at formation}\label{mformation}
The previous subsection studied halo formation, where formation was defined as the first time that the mass of one of the progenitors exceeds half the total.  Therefore, this mass can have any value between 1/2 and 1 times the final mass, and one can study the distribution of masses at, and just prior to, formation.  The excursion set constant barrier model makes a prediction for this distribution \citep{nusser}.  The mass distribution at formation is expected to be 
\begin{equation}
\label{pu}
 p(\mu)\, \rm{d}\mu = \frac{2}{\pi} \sqrt{\frac{1-\mu}{2\mu-1}}\,
  \frac{\rm{d}\mu}{\mu^{2}}, \,\,\, where \,1/2 \leq \mu \leq 1,
\end{equation}
and $\mu \equiv m/M$, and the distribution just before formation is 
\begin{equation}
\label{qu}
 q(\mu)\, \rm{d}\mu = \frac{1}{\pi(1-\mu)} 
  \Big{(} \sqrt{\frac{\mu}{1-2\mu}}-\sqrt{1-2\mu}   \Big{)} \,
  \frac{\rm{d}\mu}{\mu^{2}}, 
\end{equation}
where $1/4 \leq \mu \leq 1/2$.  We have found that, to a very good 
approximation, $\mu\,p(\mu) \to \mu\,q(\mu)$ if one replaces $\mu \rightarrow 1/4\mu$ (solid and dashed curves in Figure~\ref{mF}, 
respectively).  Let $m_B$ and $m_A$ denote the masses before and after formation, respectively.  Roughly speaking, the symmetry about $M/2$ indicates that having a specific ratio of $m_B$ to $M/2$ before formation is equaly likely to having the same ratio of $M/2$ to $m_A$ after formation. 

Although these expressions were derived assuming a white-noise power spectrum, they are expected to be relatively independent of $P(k)$.  \cite{st04} showed that they did indeed match numerical simulations well for different cosmologies and initial power spectra.  Figure~\ref{mF} shows that they also work well for square-root barriers.

The agreement between the theory curves (smooth curves) and our Monte Carlo trees (jagged lines, labeled `Cont') is excellent.  All panels show that agreement with simulation data is also quite good.  However, there is a systematic discrepancy:  the cusp at $\mu=1/2$ appears to be less pronounced in the simulation, with correspondingly lower tails.  A similar discrepancy was seen by \cite{st04}, who suggested that the fact that the simulations only provide discrete snapshots in time may be smoothing out the peak.  By sampling our trees at the simulation snapshots (open symbols, labeled `Snap'), we have attempted to model the magnitude of this effect.  Figure~\ref{zoom} illustrates that the cusp has indeed been smoothed, but this is a dramatic effect only around $0.49\le \mu\le 0.51$.  The discrepancies further from the peak remain (Figure~\ref{mF}).  

%%%%%%%%%%%%%%%%%%%%%%%%%%%%%%%%%%%%%
%   REDSHIFT OF LAST MAJOR MERGER   %
%%%%%%%%%%%%%%%%%%%%%%%%%%%%%%%%%%%%%
\subsection{The redshift distribution of the last major merger}
\label{lastmerger}

Mergers of galaxies with similar masses are expected to produce strong short-lived periods of star formation: \emph{starbursts} \citep{hern}.  Recent numerical studies suggest that the mass ratio of the galaxies involved plays an important role:  merger-induced bursts occur when the galaxies have similar masses \citep{gao04a,sh05,cox}.  Moreover, as suggested by \cite{maller}, a galaxy's Hubble type is strongly correlated with its last major merger.

Understanding such mergers requires understanding the mergers their host haloes undergo.  Consider a merger $(m',m'-m) \rightarrow m$, with $m'>m-m'$.  For ease of comparison with \cite{pch07}, we will define a `major' merger as one in which $(m-m')/m' \geq 1/3$.  The filled circles in Figure~\ref{zlmm} show the redshift distribution of the last major merger on to the main branch.  (The last major merger does not necessarily happen on the main branch.  However, Figure~3 of \cite{pch07} suggests that, in most cases, it does.  Presumably, this is because the assembly of haloes in recent times is dominated by mergers.)  Curves show measurements in our full trees (`Cont'), and open symbols show the result of only sampling the trees at the GIF2 simulation outputs (`Snap').  For the discretely-sampled data, only mergers involving haloes with at least ten particles are considered.  Note that the anomalously low data point that is second from the left in all panels appears to be an effect of seeing the tree at discrete snapshots -- the smooth curves show no such dip.  This feature is also present in the simulation analysed by \cite{pch07}; we expect it to disappear if more finely spaced snapshots are analysed.  

For high masses, the data from the square-root trees peak at about the same redshifts as the simulations; the constant barrier, spherical collapse trees peak at lower redshifts.  This improvement relative to the spherical collapse case is similar to that in the modified \textsc{galform} trees of \cite{pch07}.  However, our square-root trees tend to lie above the simulation at low redshifts, and below at higher redshifts.  The discrepancy with simulation becomes increasingly worse at small masses, although it is possible that the GIF2 results for our two smallest mass bins are not reliable -- the high redshift mergers involve haloes with few particles.  Because we require haloes to have more than 10 particles, we are likely to miss major mergers once the typical mass becomes of this order.

%%%%%%%%%%%%%%%%%%%%%%%%%%%%%%%%
%     D I S C U S S I O N      %
%     C O N C L U S I O N      %
%%%%%%%%%%%%%%%%%%%%%%%%%%%%%%%%
\section{Discussion}\label{conclusions}
We presented an algorithm for generating merger histories of dark matter haloes.  This algorithm is based on the excursion set approach (Figures~\ref{chist}, \ref{shist} and related discussion), and can handle moving barriers of the sort that are associated with ellipsoidal collapse (equation~\ref{ellbarrier}).  We illustrated its use by generating the forest of trees associated with a square root barrier (equation~\ref{sqrtB}).  The halo mass function associated with this barrier is known to provide a reasonable description of halo abundances in the GIF2 simulation against which we test our model \citep{mr05,ms08}.

Our algorithm produces merger histories which yield the progenitor mass functions which are commonly computed using excursion set theory -- demonstrating that our approach is nearly self-consistent -- over a broad range of masses and redshifts (Figures~\ref{prog1}-\ref{prog5}).  These progenitor mass functions are also in reasonable agreement with measurements in simulations; they are a significant improvement on trees based on the constant barrier, spherical collapse model.  

Our algorithm is different from others in the literature because it is based on taking discrete steps in mass, whereas all others (full N-body simulations included) take discrete steps in time.  We also used our algorithm to show that while the distribution of times at which haloes assemble half their mass depends quite strongly on the barrier shape (Figure~\ref{zF}), the distribution of masses at formation does not (Figure~\ref{mF}).  Excursion set related formulae for this `universal' distribution (equations~\ref{pu} and~\ref{qu}) provide an excellent description of our merger trees; the agreement with simulations is good, but not perfect.  

The algorithm is described in detail in Appendix~\ref{algorithm}.  In essence, it requires that one be able to generate one-dimensional random walks quickly.  Since this reduces to being able to generate long strings of Gaussian variables, and fast routines for this exist, it is reasonably fast.  Significant speed-ups can be obtained if one exploits known properties of random walks.  For example, in the present context, all steps in the walk which lie below the current threshold value are not interesting (e.g. gray-jagged portions of the random walks during the $S_m \rightarrow S_{m'}$ jump in Figures~\ref{chist}-\ref{shist}).  If the distribution of the number of steps it takes to first exceed the current value is known, then one need not generate all these steps, one can instead draw a number from this distribution, and simply jump this number of steps.  Incorporating such changes into our algorithm is the subject of ongoing work.

The excursion-set approach successfully describes many properties of the hierarchical growth of dark matter haloes.  However, with the exception of white-noise initial conditions, the progenitor distributions it predicts cannot be made consistent with the intuitively attractive notion that, in sufficiently small time-steps, mergers are binary \citep{sp97,benson,eyal2,jun,benson2}.  This accounts for some of the discrepancy between our trees and the excursion set based theory curves.  A number of authors have attempted to alleviate this by relaxing the assumption of binary mergers in their merger tree algorithms.  E.g., one of the two algorithms in \cite{sl99} reproduces the spherical collapse based results exactly, for arbitrary power spectra.  In this algorithm, mergers occur between groups of objects rather than just two but, as they pointed out, this was a rather contrived solution.  \cite{sk99}, \cite{eyal2} and \cite{zmf08} discuss other possibilities.  We make no attempt to address this issue with our algorithm. 

The appendix also shows that building the main branch is straightforward (Figure~\ref{split}).  We expect our approach to facilitate studies involving the mass accretion history (MAH) of a halo \citep{avila,nusser,vdb02,wechsler,gao05,vladimir,kyle,getal08,lidia}.  This is the subject of work in progress.

We mentioned in the introduction that halo merger histories play an important role in galaxy formation models.  Models of the mergers of supermassive black holes \citep{menouetal,vetal03,yoo,vetal08}, merger-induced starbursts \citep{hern} and quasars \citep{kh00} also have the assembly of haloes as their backbone.  Halo assembly histories also play a key role in studies of the brightest cluster galaxies \citep{delucia}, luminous red galaxies \citep{almeidaetal,conroy,masjedi,wake}, satellites and the intracluster light \citep{cwk,ssm07}, and the nature of substructure in galaxy clusters.  This last is important for interpreting the Sunyaev-Zel'dovich \citep{holderetal}, and strong lensing \citep{natetal} signals.  
  
One of the advantages of using Monte Carlo merger trees is that one may easily change the underlying cosmology and initial power spectrum.  Recently there have been attempts to describe spherical collapse and non-linear growth in modified-gravity theories \citep{schafer,laszlo}.  In principle, such modifications can easily be incorporated into our algorithm.  We hope that our algorithm will prove useful in some of these studies.

%%%%%%%%%%%%%%%%%%%%%%%%%%%%%%%%%%%
%  A C K N O W L E D G M E N T S  %
%%%%%%%%%%%%%%%%%%%%%%%%%%%%%%%%%%%
\section*{acknowledgments}
We wish to thank E. Neistein, K. Stewart, V. Avila-Reese, J. Pan, M. Martino, J. Newman and J. Hyde for useful discussions and positive feedback.  We also thank our referee, O. Fakhouri, for his insightful criticisms and genuine interest in improving this work.  J. Moreno would like to thank V. Moreno and the kids for their patience and understanding during the completion of this project.  This work was supported in part by CONACyT--Mexico, and by Grant 2002352 from the US-Israel BSF.

%%%%%%%%%%%%%%%%%%%%%%%%%%%%%%%%%%%%%%%%
%          R E F E R E N C E S         %
%%%%%%%%%%%%%%%%%%%%%%%%%%%%%%%%%%%%%%%%

\bibliographystyle{mn2e}

%%%%%%%%%%%%%%%%%%%%%%%%%%%%%%%%%%%%%%%%
%           A P P E N D I X            %
%%%%%%%%%%%%%%%%%%%%%%%%%%%%%%%%%%%%%%%%
\appendix

%%%%%%%%%%%%%%%%%%%%%%%%%%%
%    A L G O R I T H M    %
%%%%%%%%%%%%%%%%%%%%%%%%%%%
\section{The algorithm}\label{algorithm}

Consider a dark matter halo of mass $M_0$ at redshift $Z_0$ (Figure~\ref{mht}).  Realisations of its merger history may be constructed with our tree algorithm.  Below we explain how each branch is built and how these branches are connected.  

%%%%%%%%%%%%
%  BRANCH  %
%%%%%%%%%%%%
\subsection{The branch: from random walks to mass histories}\label{branchalg}

The mass growth history of a halo is contained in a random walk (Section~\ref{masshist}).  First we discuss the constant-barrier (spherical collapse) model and then the square-root barrier (ellipsoidal collapse) case.   

A random walk is essentially a collection of steps and heights:  $\{s_0, \dots , s_i, \dots \}$ and $\{h_0, \dots , h_i, \dots \}$ (e.g., the jagged line in Figure~\ref{chist}).  Consider a horizontal barrier of height $\delta_{\rm c}(z)$.  This line may intersect the walk at several values of $S$.  The smallest of such of values, $S_m$, indicates what mass the halo had at redshift $z$.  As $z$ increases, so does $\delta_{\rm c}$ -- and $m$ decreases accordingly, as expected in any hierarchical model of halo assembly.  

As we increase the height of the barrier, a subset $\{(S_0,H_0), \dots , (S_j,H_j), \dots \}$ of the walk is chosen.  These points contain the mass history (e.g., the dark-filled circles in Figure~\ref{chist}).  Every point in this subset has the following property: they are higher than all their predecessors.  This is illustrated by the point at $(S_m,\delta_{\rm c}(z))$ Figure~\ref{chist}: it has the maximum height in the range $S_M \leq S \leq S_m$.  In other words, for any point $(s_i,h_i)$ on the walk to be selected as part of the history, it must satisfy the following condition:
\begin{equation}
\label{histreq}
h_i>h_k,\,\,\forall \, k<i. 
\end{equation}
Now we discuss what modifications are necessary when the square-root barrier model of ellipsoidal collapse is used (Figure~\ref{shist}).  Consider a barrier (equation~\ref{sqrtB}) with $\delta$-intercept $\sqrt{q}\delta_{c}(z)$.  This curve may intersect the random walk at several places.  We pick the smallest of these to find the mass at that redshift.  As $z$ increases, the $\delta$-intercept increases, but the overall shape of the barrier remains unchanged.  This is how the mass history $\{(S_0,H_0), \dots , (S_j,H_j), \dots \}$ is selected.  From this subset we can
construct a string $\{\sqrt{q}D_0, \dots , \sqrt{q}D_i, \dots \}$ of `$\delta$-intercepts', where
\begin{equation}
D_i=\frac{1}{\sqrt{q}}(H_i-\beta\sqrt{S_i})\footnote{Note that this mapping is ill-defined when $\gamma>1/2$, because these barriers cross.}
\end{equation}
Every point on the history has the following property: its $\delta$-intercept is higher than that of its predecessors.  In other words, for a point $(s_i,h_i)$ on the walk to belong to the mass history, condition~(\ref{histreq}) is replaced by
\begin{equation}
d_i>d_k,\,\,\forall \, k<i, 
\end{equation}
where
\begin{equation}
d_i=\frac{1}{\sqrt{q}}(h_i-\beta\sqrt{s_i}).
\end{equation}
For a given cosmology and initial power spectrum one may map excursion set variables into physical ones: $(S(m),\delta_{\rm c}(z)) \leftrightarrow (m,z)$.  With this prescription, the mass accretion history is obtained:  $\{(M_0,Z_0), \dots , (M_j,Z_j) \dots \}$.  The masses in the history are such that $S(M_j)=S_j$ and the redshifts are given by $\delta_{\rm c}(Z_j)=D_j$.  When barriers are constant, $D_j=H_j$.  

%%%%%%%%%%%%%%
%  THE TREE  %
%%%%%%%%%%%%%%
\subsection{The tree: connecting the branches}\label{treealg}

%%%%%%%%%%%%%%
\begin{figure}
\centering
\includegraphics[width=0.9\columnwidth]{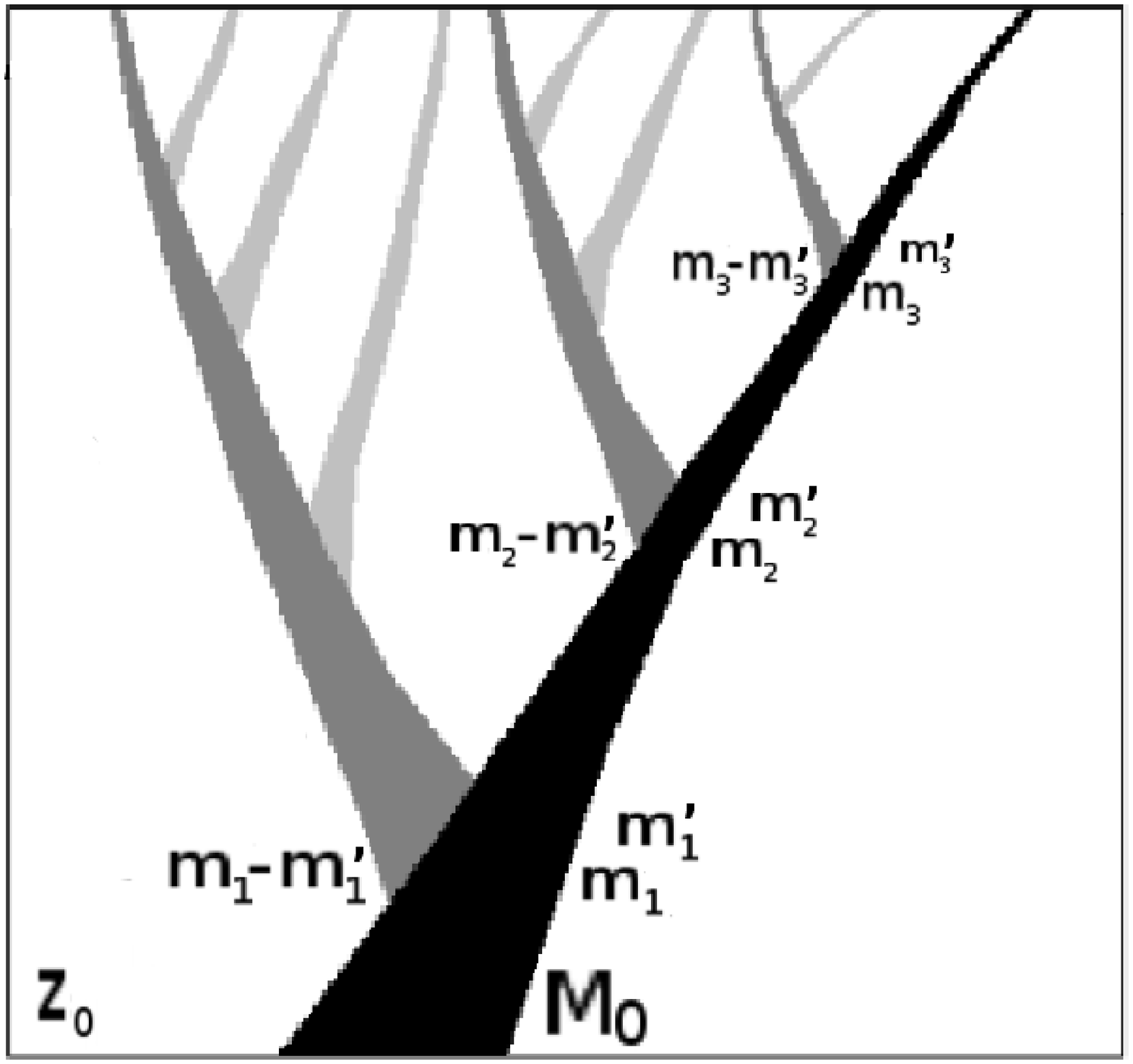}
\caption{A sample merger history tree of a halo with mass $M_0$ at $Z_0$.  The mass history (black branch) is constructed with a random walk.  Mass jumps larger than $m_{dust}$ are identified, and branches are connected there (medium-gray).  This process is repeated for each branch, and new branches are connected (light-gray) until the tree is complete.}
\label{mht}
\end{figure}
%%%%%%%%%%%%

Consider a halo of mass $M_0$ at redshift $Z_0$.  We are interested in constructing its merger history tree.  The first step is to draw a random walk with origin at 
\begin{equation}
s_0=S(M_0),\,\,\,h_0=B(S(M_0),\delta_{c}(Z_0)),
\end{equation}
where $B(S,\delta_c)$ is given by equation~(\ref{sqrtB}).  The mass history is then collected: $\{(M_0,Z_0), \dots , (M_j,Z_j), \dots \}$ (see section~\ref{branchalg}).  In Figure~\ref{mht}, this is represented by the black branch.  

The mass history can also be seen as a series of jumps in mass:  $\{(M_0 \leftarrow M_{0}',Z_0), \dots , (M_j \leftarrow M_{j}',Z_j), \dots \}$, with $M_{j}'=M_{j+1}$.  Such jumps are interpreted as binary mergers: 
$M_{j}'+(M_{j}-M_{j}') \rightarrow M_{j}$.  In practice, we only care about the mass jumps above the branching-mass resolution. Denote this subset of jumps as $\{ \dots , (m_J \leftarrow m_{J}',z_J), \dots \}$, where
\begin{equation}
\label{connect}
m_J-m_{J}'>m_{dust}.  
\end{equation}
These jumps are shown in Figure~\ref{mht}.  The next step is to construct the history of each halo with mass ($m_{J}-m_{J}'$) that falls on to the mass history of the $M_0$-halo.  This is done by generating random walks originating from
\begin{equation}
s_0=S(m_J-m_{J}'),\,\,\,h_0=B(S(m_J-m_{J}'),\delta_{c}(z_J)),
\end{equation}
Such mass histories correspond to the medium-gray branches in Figure~\ref{mht}.  The above process is repeated for each of these branches:  retrieve their mass history, identify large enough mass jumps, and attach 
new branches there (light-gray in Figure~\ref{mht}).  Eventually, all the new branches will only involve mass jumps that do not satisfy condition (\ref{connect}).  At this point the tree is complete and the algorithm stops.

%%%%%%%%%%%%%%%%%
%  MAIN BRANCH  %
%%%%%%%%%%%%%%%%%
\subsection{The main branch: a simple modification}\label{mainalg}

\begin{figure}
\centering
\includegraphics[width=0.9\columnwidth]{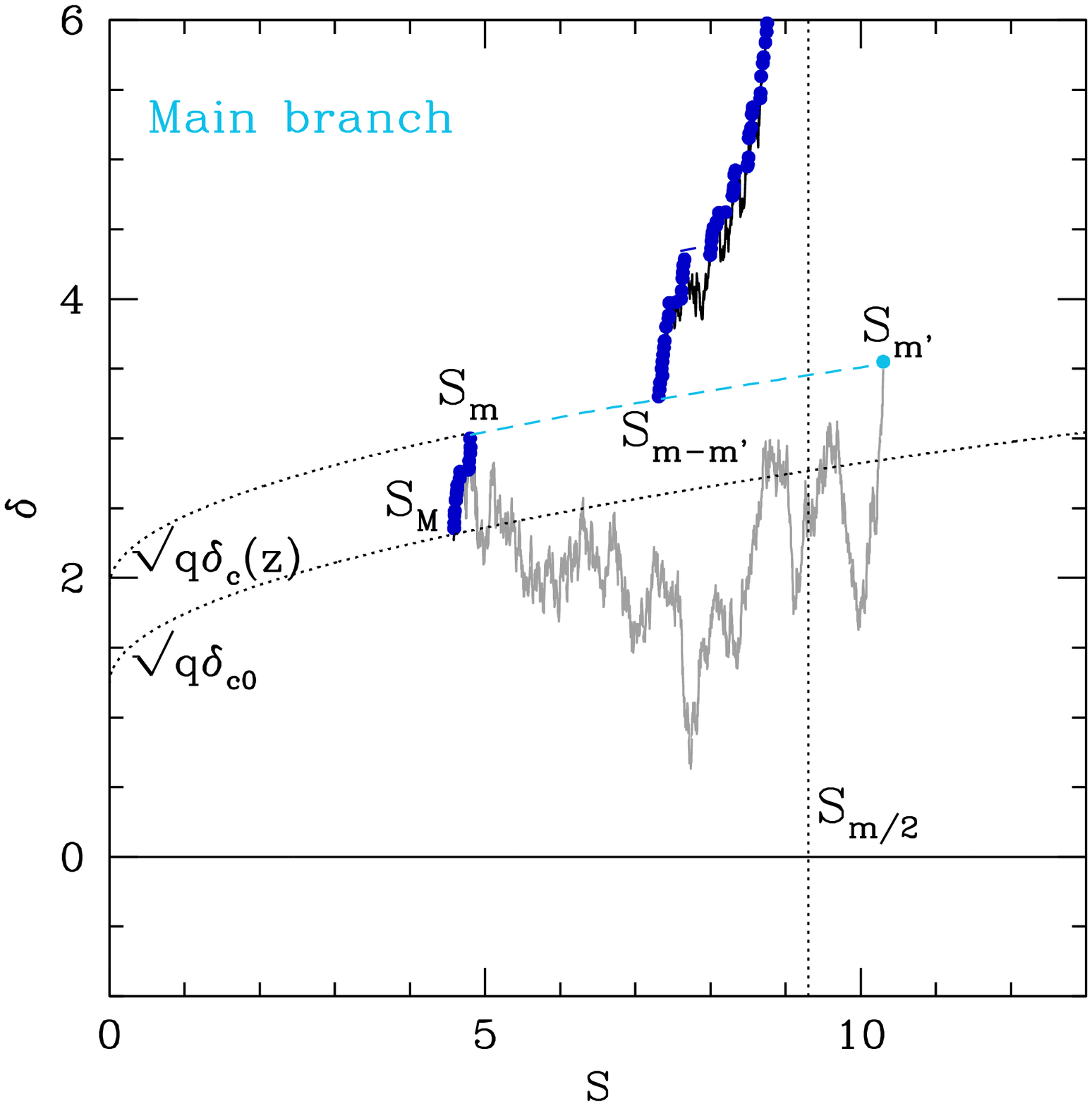}
\caption{Main branch algorithm.  The symbols and styles are the same as in Figure~\ref{shist}.  The dotted vertical line denotes $S=S_{m/2}$.  The main branch is built by following the most massive piece at each split:  $S_{m'}>S_{m/2}$ indicates that $m'<m/2$, implying that $m-m'>m'$, whose mass history we must follow.}
\label{split}
\end{figure}

In a merger history tree, the `main' branch of a halo is obtained by following the most massive progenitor at each mass split.  In this section, we show that the mass history algorithm described in section~\ref{branchalg} can be easily modified to construct the main branch.  This process requires continuous monitoring of the walk and its associated history, and is illustrated in Figure~\ref{split} (compare to Figure~\ref{shist}).   

Recall that mass decreases with increasing $S$.  For the portions in the mass history consisting of jumps where the mass loss is less than half, the algorithm is unchanged (e.g., the portion with $S_M \leq S \leq S_m$ in Figure~\ref{split}).  Occasionally, there are jumps where more than half the mass is lost.  Such is the case for the $S_m \rightarrow S_{m'}$ jump illustrated in Figure~\ref{split}, with $S_{m'}>S_{m/2}$ (i.e., $m'<m/2$ and $(m-m')>m/2$).  To construct the main branch in that situation, one must simply follows $(m-m')$, not $m'$.  In other words, instead of continuing the walk at $(S_{m'},\sqrt{q}\delta_{\rm c}(z)+\beta\sqrt{S_{m'}})$, one must continue from $(S_{m-m'},\sqrt{q}\delta_{\rm c}(z)+\beta\sqrt{S_{m-m'}})$ (i.e., the dark filled circles in Figure~\ref{split}).

%%%%%%%%%%%%%%
%%%%%%%%%%%%

\label{lastpage}
\end{document}